\documentstyle[draft]{mn}

\title[Moment Analysis Applied to LMC Star Clusters]
      {Moment Analysis Applied to LMC Star Clusters}

\author[T. Banks, R.J. Dodd and D.J. Sullivan]
       {Timothy Banks$^{1}$, R.J. Dodd$^{2}$, and D.J. Sullivan$^{1}$\\
	$^{1}$Department of Physics, Victoria University of Wellington,
	P.O. Box 600, Wellington, New Zealand. \\
	$^{2}$Carter Observatory, P.O. Box 2909, Wellington, New Zealand. }
\date{}
\pagerange{\pageref{firstpage}--\pageref{lastpage}}

\def\LaTeX{L\kern-.36em\raise.3ex\hbox{a}\kern-.15em
    T\kern-.1667em\lower.7ex\hbox{E}\kern-.125emX}

\begin{document}
\label{firstpage}

\maketitle

%--------------------extra commands section begins-------------------------

%       \sun : sun symbol               \fd : fraction of a day
%       \diameter : diameter symbol     \fm : fraction of a minute
%       \degr : degree                  \fh : fraction of an hour
%       \arcmin : guess what            \fs : fraction of a second
%       \arcsec : ditto                 \fp : fraction of a period
%       \farcs : fraction of arcsecond  \farcm : fraction of arcmin

\tolerance=10000

\def\diameter{{\ifmmode\mathchoice
{\ooalign{\hfil\hbox{$\displaystyle/$}\hfil\crcr
{\hbox{$\displaystyle\mathchar"20D$}}}}
{\ooalign{\hfil\hbox{$\textstyle/$}\hfil\crcr
{\hbox{$\textstyle\mathchar"20D$}}}}
{\ooalign{\hfil\hbox{$\scriptstyle/$}\hfil\crcr
{\hbox{$\scriptstyle\mathchar"20D$}}}}
{\ooalign{\hfil\hbox{$\scriptscriptstyle/$}\hfil\crcr
{\hbox{$\scriptscriptstyle\mathchar"20D$}}}}
\else{\ooalign{\hfil/\hfil\crcr\mathhexbox20D}}%
\fi}}

\def\sun{\hbox{$\odot$}}
\def\degr{\hbox{$^\circ$}}
\def\arcmin{\hbox{$^\prime$}}
\def\arcsec{\hbox{$^{\prime\prime}$}}
\def\utw{\smash{\rlap{\lower5pt\hbox{$\sim$}}}}
\def\udtw{\smash{\rlap{\lower6pt\hbox{$\approx$}}}}
\def\fd{\hbox{$.\!\!^{\rm d}$}}
\def\fh{\hbox{$.\!\!^{\rm h}$}}
\def\fm{\hbox{$.\!\!^{\rm m}$}}
\def\fs{\hbox{$.\!\!^{\rm s}$}}
\def\fdg{\hbox{$.\!\!^\circ$}}
\def\farcm{\hbox{$.\mkern-4mu^\prime$}}
\def\farcs{\hbox{$.\!\!^{\prime\prime}$}}
\def\fp{\hbox{$.\!\!^{\scriptscriptstyle\rm p}$}}
\def\loa{\mathrel{\mathchoice {\vcenter{\offinterlineskip\halign{\hfil
$\displaystyle##$\hfil\cr<\cr\approx\cr}}}
{\vcenter{\offinterlineskip\halign{\hfil$\textstyle##$\hfil\cr
<\cr\approx\cr}}}
{\vcenter{\offinterlineskip\halign{\hfil$\scriptstyle##$\hfil\cr
<\cr\approx\cr}}}
{\vcenter{\offinterlineskip\halign{\hfil$\scriptscriptstyle##$\hfil\cr
<\cr\approx\cr}}}}}

%---------------------------end of extra commands----------------------------


\begin{abstract}
Statistical moment-based ellipse fitting was performed on observations of Large
Magellanic
Cloud clusters, confirming that trends are evident in their position angles and
ellipticities, as had been reported in the literature.
Artificial cluster images with known parameters were generated, and
subjected to the same analysis techniques, revealing apparent trends caused by
stochastic processes. Caution should therefore be exercised in the
interpretation
of observational trends in young LMC clusters.
\end{abstract}

\begin{keywords}
Star Clusters -- Ellipticity -- Large Magellanic Cloud
\end{keywords}


\section{Introduction}

Globular clusters belonging to the Galaxy are essentially spherical
in shape, with a mean ellipticity (defined as for elliptical
galaxies) of 0.12 (Shapley, 1930). One of the most elliptical galactic
globular clusters is $\omega$ Centauri with  ellipticity estimates of 0.14
(Dickens and Woolley, 1967) and 0.19 (Van den Bergh, 1983; Frenk \&
Fall, 1983).

It has been known for many years that some of the brightest
old clusters in the Magellanic Clouds are strongly flattened (Van den
Bergh, 1983). Geisler \& Hodge (1980) used microdensitometry of
photographic plates for 25 Large Magellanic Cloud (LMC) star clusters, and
found a mean ellipticity of 0.22, which they commented was
far above the galactic mean. They also noted some internal variations
in the position angle of the fitted ellipses as well as in the ellipticity
itself.  Radial variations were reported by
Geyer, Hopp \& Nelles (1983) and Zepka \& Dottori (1987),
and commented on by Kontizas {\ et al.} (1989) as a possible explanation
for the discrepancies in the elliptical parameters derived by different
investigators for the same clusters (see Table \ref{table:twod}).
Van den Bergh (1983) noted that the
more luminous LMC clusters of any age are more flattened than fainter
clusters. Frenk \& Fall (1982) estimated
by eye 52 LMC cluster ellipticities, although their mean value was 0.12
$ \pm $ 0.07, which should be compared with their mean estimate for 93
galactic globulars of 0.08 $ \pm $ 0.05. Using the age classes of
Searle, Wilkinson \& Bagnuolo
(1980), Frenk \& Fall noted an apparent 97~\% correlation of
ellipticity with age, with the younger clusters being flatter on average
than the older clusters which were similar to galactic globulars in shape.
They suggested this was the result of internal evolution of the clusters.
Van den Bergh~\& Morbey (1984) demonstrated that this correlation was not
statistically significant after the dependency of ellipticity on
luminosity was removed, given that the brightest blue clusters are more
luminous than the brightest red clusters, so making these two correlations
not mutually exclusive. A two-tailed Kolmogorov-Smirnov test showed that
the hypothesis that the LMC  and galactic globulars are from the same
parent population with the same ellipticity distribution can be rejected
at the 99.2\% confidence level.

The aim of the current paper was to investigate the internal variations
of elliptical parameters in a sample of LMC clusters, observations of which
were collected at the Mount John University Observatory (MJUO), New Zealand.
Due to the small (1m) size of the telescope and the primary aim of the
campaign being the derivation of colour magnitude diagrams,
the observed clusters tend to be young (Searle { et al.} (1980)
Class III and lower). Initially, a standard package was used to fit
ellipses to the clusters, but
dissatisfaction with its results led to the use of a moments based
technique (as outlined by Stobie 1980a,b and described below).

\begin{table}
\centering
\begin{tabular}{||l|c|c||c|c||c||c||}
\hline
NGC & \multicolumn{2}{c||}{F \& F} & \multicolumn{2}{c||}{G \& H}
& K  & G  \\
 & Ellip & PA & Ellip & PA &  Ellip &  Ellip \\
\hline
1651 & 0.03 & - & 0.31 & 126 & - & - \\
1751 & 0.12 & 55 & 0.26 & 133 & 0.12:0.9 & - \\
1751 &  -   & -  &  -   &  -  & 0.16:0.5 & - \\
1754 & 0.06 & 38 & 0.11 & 108 & - & - \\
1783 & 0.25 & 64 & 0.19 & 72 & - & \\
1786 & 0.02 & - & 0.09 & 166 & 0.12 & 0.00 \\
1806 & 0.09 & 9 & 0.19 & 159 & 0.07:1.8 & 0.05 \\
1806 &  -   & -  &  -   &  -  & 0.12:0.9 & - \\
1835 & 0.21 & 88 & 0.17 & 77 & 0.21:0.16 & 0.21 \\
1835  &  -   & -  &  -   &  -  & 0.14:0.13 & - \\
1835  &  -   & -  &  -   &  -  & 0.19:0.3 & - \\
1846 & 0.08 & 151 & 0.23 & 129 & 0.07:1.9 & 0.13 \\
1846  &  -   & -  &  -   &  -  & 0.16:0.11 & - \\
1917 & 0.07 & 168 & 0.29 & 4 & 0.15 & 0.00 \\
1978 & 0.33 & 150 & 0.30 & 159 & - & \\
2019 & 0.07 & 139 & 0.20 & 124 & 0.17:1.0 & - \\
2019  &  -   & -  &  -   &  -  & 0.23:0.5 & - \\
2108 & 0.11 & 80 & 0.18 & 115 & - & - \\
2121 & 0.18 & 21 & 0.32 & 64 & - & - \\
2154 & 0.17 & 36 & 0.13 & 42 & - & - \\
2155 & 0.08 & 102 & 0.27 & 76 & - & - \\
2173 & 0.06 & 70 & 0.27 & 141 & - & - \\
2210 & 0.07 & 78 & 0.12 & 82 & - & - \\
2213 & 0.01 & - & 0.26 & 91 & - & - \\
\hline
\end{tabular}
\caption{ { Literature values for ellipticities and position angles}
(PAs) of LMC clusters common to Frenk~\& Fall (1982) and \label{table:twod}
Geisler~\& Hodge (1980). The latter is given in the column G~\& H, and
the former under F~\& F.
 Where available, the ellipticity values
of Kontizas~{ et al.} (1989) and Geyer~{et al.} (1983) are also
shown. These are in columns headed K  and G respectively.
Note the frequent discrepancies, even for clusters which are
given as very elliptical in one catalogue (e.g. NGC 2121, 2155, 2173
etc). Kontizas~{ et al.} claim that such differences are due to
internal variations in the clusters, and so the radii that an ellipticity
is measured at must also be specified. Where Kontizas~{ et al.} have given
more than one value for a cluster, the values are given on subsequent
lines. The radius the ellipticity was measured at is given in arcminutes
as the value after the colon. Generally they
measured ellipticity at the half mass radius (see King, 1966a), which is
a constant throughout a cluster's dynamical evolution.
}
\end{table}


\section{Method}

A major advantage in using moments (see Larson, 1982, for a general
background) for image analysis lies in their
ease of calculation, e.g. the summation for the second moment can
be calculated without knowledge of the mean, which is only needed in
the final step:

\[
\sum_{i \: = \: 1}^{n}  ( \: x_i \: - \: \bar{x} \: )^2 \: f_i\: =
\:  \sum_{i \: = \: 1}^{n} x_i^2 f_i \: - \:
\bar{x}^2 \sum_{i \: = \: 1}^{n} f_i
\]

where $ f_i $ is the intensity of the $ i $ -th background
subtracted $ x $ pixel.
The $ x $ axis zeroth and normalised first order moments

\[
\sum_{i \: = \: 1}^{n} f_i \: \: {\rm and }
 \: \: \frac{\sum_{i \: = \: 1}^{n}
x_i f_i } { \sum_{i \: = \: 1}^{n} f_i }
\]

can be easily interpreted as the object's total intensity
and the normalised intensity weighted
$ x $ centroid. The second order moments are harder to
interpret, but give information on the structure of the pixel cluster.
For a continuous two dimensional distribution, the second order moments are:

\[
U_{\rm xx} \: = \: \int \! \! \int \frac {( \: x_1 \: - \: \bar{x_1} \: ) ^ 2
\: dx_1 \: dy_1 } { A }
\: {\rm , } \:
\]
\[
U_{\rm yy} \: = \:
\int \! \! \int \frac  {( \: y_1 \: - \: \bar{y_1} \: ) ^ 2 \: dx_1 \: dy_1 } {
A}
\:
{\rm , \: and } \:
\]
\[
U_{\rm xy} \: = \:
\int \! \! \int \frac  {( \: y_1 \: - \: \bar{y_1} \: ) ( \: x_1 \: - \:
\bar{x_1} \:
)  \: dx_1 \: dy_1 } { A }
\]
where area $ A $ of the ellipse
is $ {\rm \pi } a b $. $a$ and $b$ are defined conventionally as the ellipse
semi-axes.
We assume that the ellipse is centred
on the origin, i.e $ \bar{x_1} $ = $ \bar{y_1} $ = 0.
Following Stobie (1980a,b) the size, orientation ($ \theta $), and ellipticity
($ e \: = \: 1 \: - \: \frac{b}{a}$) can be calculated of an ellipse fitted to
a distribution above the detection level:
\begin{equation}
\theta \: = \: \frac { 1 } { 2 } \arctan \left(
\frac { 2 U_{\rm xy} } { U_{\rm xx} \: - \: U_{\rm yy} } \right)
\label{eq:sau}
\end{equation}

\begin{equation}
a \: = \: \sqrt { 2 \left ( ( U_{\rm xx} \: + \: U_{\rm yy} ) +
\sqrt { \left ( ( U_{\rm xx} \: - \: U_{\rm yy} ) ^ 2 \: + \: 4 U_{\rm xy}^2
\right) }
\: \right)}
\label{eq:bay}
\end{equation}

\begin{equation}
\label{eq:tam}
b \: = \: \sqrt { 2 \left( ( U_{\rm xx} \: + \: U_{\rm yy} ) \: - \: \sqrt {
\left(
( U_{\rm xx} \: - \: U_{\rm yy} ) ^ 2 \: + \: 4 U_{\rm xy}^2 \right)} \:
\right) }
\end{equation}

Similar equations can be derived for weighted moments.
Weighted first order moments were used to determine the centre of the pixel
distribution (Dodd \& MacGillivray, 1986), while
the ellipse
fitting applied no weighting. The latter results in the
distribution's determined shape and orientation being more representative
of the overall distribution, rather than being skewed by the brighter
central regions. In asymmetric distributions, the former point means that
the derived centre is the centre of mass.


\subsection{The fitting software}
A Fortran program was written based on equations
\ref{eq:sau}, \ref{eq:bay}, and \ref{eq:tam}, using the
IRAF\footnote{Image Reduction and Analysis Facility:
courtesy of the National Optical Astronomical Observatories,
which are operated by the Association of Universities for Research in
Astronomy under co-operative agreement with the National Science
foundation.} Imfort libraries for the image manipulation routines (details
on IRAF at VUW may be found in Banks, 1993). Threshold values were read
from a file. The image's background was estimated using the IRAF {\it
imex} tool in clear (star and cosmic strike free) regions of the image.

The image pixel array was scanned starting at
(1,1). The array dimensions were automatically determined by the software.
The first direction of search was along the x axis. When the end
of this row was reached, the x position was reset to 1 and the y position
incremented. This continued until the opposite corner of the image was met.
When the intensity of the search's current pixel was
above the threshold set for the ellipse fitting, an interior defined
seed fill algorithm (see p86, Rogers, 1988) was commenced, and
the x and y positions of the pixel were pushed onto a stack. If the stack
contained
other pixel values, the following occurred: The pixel positions were popped
from the stack, and the corresponding (x,y) position flagged
in a boolean array of identical dimensions to the image. Eight way
connectivity (see p84, Rogers, 1988)
was assumed, so all pixels around the current one were examined in
turn. If the new pixel had an intensity above the threshold and had
not been marked as detected, its position was pushed on to the stack.
The end result was that all pixels with intensities above the threshold
and contiguous were detected as an individual pixel distribution. This
subarray was then passed to the moments analysis subroutine,
to evaluate a, b, and ${\rm \theta }$ . The size, centre,
orientation, and ellipticity of the distribution
were then written to disk. Once all pixel distributions had been detected and
measured, the next value in a file containing the intensity threshold
values was read. The detection array was cleared, and the search commenced
again from position (1,1).  In practice, the threshold
step direction was towards the background, corresponding to an increase
in the dimensions of the pixel distribution, allowing examination of the
variability of the ellipse parameters describing the star cluster with
radius from the cluster's centre. Further software allowed the preparation
of isophotal maps of the ellipses, as well as graphs of the detected
pixels for a given threshold.

\begin{table}
\begin{center}
\begin{tabular}{||c|c|c|c|c|||}
\hline
\multicolumn{3}{||c|}{Ellipticity $e$} & \multicolumn{2}{c||}{Angle $\theta$}
\\
In & Out & Ratio & In & Out \\
\hline
0.033 & 0.034 & 103.0 & 90.0 & 89.5 \\
0.200 & 0.199 & 99.7 & 0.0 & 0.0 \\
0.250 & 0.250 & 100.0 & 0.0 & 0.0 \\
0.500 & 0.495 & 98.9 & 115.0 & 114.5 \\
0.571 & 0.569 & 99.3 & 110.0 & 109.9 \\
0.631 & 0.626 & 99.2 & 45.0 & 45.0 \\
0.792 & 0.784 & 98.9 & 145.0 & 145.0 \\
0.800 & 0.794 & 99.3 & 175.0 & 175.1 \\
\hline
\end{tabular}
\end{center}
\caption{ { Ellipticity and Orientation Angle
values:} \label{table:dodd} Input and output
ellipticities ($e$) and orientation angles ($\theta$)
are shown for a few selected tests, showing that the input values
are well recovered by the moment analysis technique. Ratio gives the output
ellipticity
as a percentage of the input value. The angles are given in degrees. }
\end{table}

This program was carefully tested. Initially uniformly weighted ellipses
were generated and placed into an array with an arbitrary background level.
These were then submitted as noiseless images to the program. 757 trials
with different input $ a $, $ b $, and $ \theta $ values showed that the
$ a $ and $ b $ were generally slightly overestimated by $ \sim $ 0.08 $\pm$
0.06
pixels (see Table~\ref{table:dodd} for
some representative tests). The variations with angle are
due to the discrete nature
of the fitted ellipse as pixels. As the ellipticity of the cluster
became less in tests,
it was found that although the scatter was constant, the difference for $ a $
decreased while that for $ b $ increased, as might be expected.
The ellipse generating function of the testing program
used the parametric equations for an ellipse to define the boundary,
and then block-filled the interior. Real numbers for the boundary's
x and y positions were converted into integers, indicating a boundary
pixel. In comparison with the superior approach of recognizing that pixel
($ x $, $ y$)
extends $\rm \pm 0.5$ in both dimensions, the x and y centroids are displaced
by -0.5
from the input values.

The typical FWHM of the programme images was 5 pixels and the ellipse fitted to
the star
clusters ranged out to 100 pixels. The trials showed that $\theta$ was
recovered by the
program to within 0.10 $\pm$ 0.27 degrees and the ellipticity to 0.006 $\pm$
0.003.


\section{Observations}

The observations used were collected over the 1991/92 southern hemisphere
summer with
the 1m McLellan telescope at MJUO ($\rm170^o$ 27.9 East,
$\rm43^o$ $ 59^{\rm'}.2$ South) in typically $ \sim3 $ arcsec seeing,
using a cryogenically cooled
Thomson TH7882 CDA charge-coupled device.  This chip has
384 by 576 pixels, each 23 $\mu$\/m square,
which at the f/7.9 cassegrain focus used by this study
corresponds to 0.60 arcsec
{(Tobin, 1989)}. Images were collected
using the Photometrics PM-3000 computer running FORTH
{(Moore, 1974)} software with
extensive local modifications {(see Tobin, 1991)}, and written
to half inch 9 track magnetic tape for transportation back
to Victoria University of Wellington (VUW)
for analysis. Images from these tapes are
then converted into the FITS (Wells, Griesen \& Harten, 1981)
format from the native Photometrics one,
and were read into IRAF
for subsequent reduction. Details on the data pathway
and image processing facility established at VUW can be found in
Banks (1993). Further details on the Mount John data
acquisition system and its characteristics may be found in
Tobin (1992).

We initially used the STSDAS{\footnote{Space Telescope Science Data
Analysis System, courtesy of the Space Telescope Science Institute.}}
{\it ellipse} function, based on Jedrzejewski (1987), to fit elliptical
contours to the clusters. Attempts were made to use DAOphot (Stetson,
1987) to subtract the resolvable stars which would skew the
ellipses, and to boxcar smooth these V band residuals over
an area comparable to the FWHM. Even when these steps
were taken {\it ellipse}\  often failed to find
solutions over a wide range of radii. However, successful solutions
were found
for some clusters, such as NGC 1818, which is discussed as an example.
When the brighter stars (above
a peak intensity of 300 counts) were subtracted from NGC 1818 an angle $ \theta
$
around 32 degrees was derived. The ellipticity increased from
0.070 (at a radius of 24 pixels) to a maximum of 0.280 at 33 pixels,
before dropping to $ \sim $ 0.15 at a 40 pixel radius. Frenk and
Fall (1982) give an ellipticity of 0.24 and a position angle of
115 degrees (or a $ \theta $ of 25 degrees in our notation, as $\theta$ is the
rotation
anticlockwise from a standard Cartesian x axis), in reasonable
agreement with our results. They do not give a radius for this ellipse, but
the ellipses were fitted by eye in the range between the burnt-out centres and
the peripheries defined by the background.
It should be noted that when
{\it ellipse}\ was run on the ``raw'' image of NGC 1818 a uniform
ellipticity of $ \sim $ 0.2 was found over the radius 10 to 40 pixels.

Similar problems to the NGC 1818 fits were found with NGC 1850, whose raw image
resulted
in a noisy but effectively constant ellipticity but whose star subtracted
image had a linearly increasing ellipticity with radius, and NGC 1856
which exhibited the opposite behaviour in both cases. Zepka
\& Dottori (1987) fitted ellipses to isophotal contours, and
had also noted radial variations in ellipticity
and/or axis orientation in all but 4 of their 17 LMC clusters, with
a preference for ellipticity to decrease with radius.
Fischer { et al.} (1993) used the {\it ellipse} program to fit
CCD observations of
NGC 1850. Although not numerically giving the results, they commented
that the ellipses did not fit the distribution well, that the
elliptical parameters varied rapidly with radius, and that there was
poor agreement between the B and V band fits despite no radial colour
gradient being evident in the cluster. The latter contradicts
Geyer { et al.} (1983).

\begin{table}
\centering
\begin{tabular}{||c|c|c|c||}
\hline
Cluster & Theta & Ellip & Radius \\
\hline
NGC 1818: & 0  & +   & 15-137\arcsec\ \\
NGC 1835: & 0  & +   & 17-90\arcsec\ \\
NGC 1836: & -- & --  & 3-130\arcsec\ \\
NGC 1839: & 0  & +   & 17-145\arcsec\ \\
NGC 1847: & -- & --  & 5-72\arcsec\ \\
NGC 1850: & 0  & +   & 3-117\arcsec\ \\
NGC 1856: & +  & 0   & 30-107\arcsec\ \\
NGC 2004: & 0  & +   & 7-133\arcsec\ \\
NGC 2031: & 0  & --  & 20-120\arcsec\ \\
NGC 2133: & -- & 0   & 15-125\arcsec\ \\
NGC 2164: & -- & 0   & 15-105\arcsec\ \\
NGC 2214: & 0  & 0   & 15-150\arcsec\ \\
\hline
\end{tabular}
\caption{
{ Variation of Elliptical Parameters} for selected LMC
Cluster V images. A + indicates that the parameter increased with
radius, a -- the opposite, and 0 stands for constant with radius.
The typical range of the trends were of several tens of degrees in angle
and 0.4 in ellipticity. Only general trends
are discussed, as in Zepka \& Dottori (1987). Noting that the table captions
are
reversed in Zepka \& Dottori (1987), our results agree with them for the
clusters NGC 1835, 2004, and 2214, but not for NGC 2031 (which they found to
vary in
both parameters). Trends evident in NGC 1835 were smooth.\label{table:once}
The column `Radius' gives the range of radii that ellipses were fitted over.
}
\end{table}

Concerned at this lack of reliability and unsure if any derived trends
were real, we adopted the moments technique outlined above, which appeared to
be more robust. As a trial, the first image to be fitted was of M81 obtained
by Michael Richmond (Princeton), who kindly made the image available.
Ellipse fitting to the smooth distribution of this Sb galaxy showed
constant values of around 0.3 and $ \rm 150^o $ for the ellipticity
and the position angle. If the disk of the galaxy is assumed to be circular,
then M81 is tilted to the line of sight by some 46 degrees. The
ellipticity is somewhat less than the 0.5 given in Allen (p288, 1973), as our
measure is for the inner nuclear region of the galaxy, not including the spiral
arms.
The position and inclination angles are in good agreement with
the $\rm 150^o $ and $\rm \sim 35^o $ values of Boggis (p288, Jones, 1968).

Having demonstrated
the stability of the fitting program on a smooth distribution, we then
proceeded to our own LMC cluster observations, with their more irregular
morphology partly due to the presence of bright young stars.
Table~\ref{table:once}
gives the general results of the fitting program on
images with the resolvable stars subtracted, while
Table~\ref{table:third_table} presents
a comparison for the derived ellipticities with those of Zepka~\& Dottori
(1987) and
Kontizas et al. (1989).
While overall trends were apparent, large radial variations in $\theta$ and $e$
were
present even when bright stars were removed and smoothing employed (as before).
The results given in Table~\ref{table:third_table} reveal a somewhat weak
agreement between
the three studies, which used different techniques on different
data sets collected under different seeing conditions. In the case of
the current study, the poor seeing experienced would increase the difficulty of
detecting and
removing bright stars from the clusters.
Both the previous studies used PDS scans of photographic plates.
Zepka~\& Dottori (1987) performed least squares fits to isophotal contours,
while
Kontizas et al. (1989) used a computer-aided interactive procedure where
ellipses
were fitted by eye to scanned images. No mention is made of seeing by either
study.
The ellipticities of Kontizas et al. (1989) appear to be biased
towards the values at the maximum radius.
There are major differences between Kontizas et al. (1987)
and Zepka~\& Dottori (1987) for some clusters,
as well as discrepancies with the current study.

\begin{table}
\centering
\begin{tabular}{||c|c|c|c|c||}
\hline
         & Radius   & \multicolumn{3}{c||}{Ellipticity} \\
Cluster  & (Arcsec) & BDS        & K       & ZD  \\
\hline
NGC 1835  & 27      & 0.09       &         & 0.16 $\pm$ 0.09 \\
$\cdots$  & 34      & 0.11       &         & 0.21 $\pm$ 0.12 \\
$\cdots$  & 46      & 0.12       &         & 0.26 $\pm$ 0.12 \\
$\cdots$  & 48-90   & 0.14-0.24  & 0.14    &                 \\
NGC 1847  & 18-30   & 0.48-0.62  & 0.29    &                 \\
$\cdots$  & 30-54   & 0.15-0.62  & 0.20    &                 \\
NGC 1850  & 30      & 0.20       & 0.19    &                 \\
$\cdots$  & 48-192  & 0.18-0.35  & 0.10    &                 \\
NGC 1856  & 24-84   & 0.01-0.10  & 0.16    &                 \\
$\cdots$  & 96      & 0.05       & 0.05    &                 \\
NGC 2004  & 42-60   & 0.02-0.21  & 0.20    &                 \\
$\cdots$  & 66-96   & 0.18-0.28  & 0.16    &                 \\
NGC 2031  & 12-54   & 0.21-0.44  & 0.20    &                 \\
$\cdots$  & 17      & 0.44       &         & 0.46 $\pm$ 0.09 \\
$\cdots$  & 22      & 0.30       &         & 0.28 $\pm$ 0.13 \\
$\cdots$  & 30      & 0.27       &         & 0.21 $\pm$ 0.13 \\
$\cdots$  & 51      & 0.29       &         & 0.21 $\pm$ 0.08 \\
$\cdots$  & 60-114  & 0.11-0.26  & 0.11    &                 \\
NGC 2214  & 19(a)   & 0.45       &         & 0.40 $\pm$ 0.14 \\
$\cdots$  & 19(b)   & 0.45       &         & 0.38 $\pm$ 0.12 \\
$\cdots$  & 26(b)   & 0.37       &         & 0.39 $\pm$ 0.12 \\
$\cdots$  & 27(a)   & 0.36       &         & 0.34 $\pm$ 0.13 \\
$\cdots$  & 36(b)   & 0.34       &         & 0.46 $\pm$ 0.09 \\
$\cdots$  & 43(a)   & 0.32       &         & 0.30 $\pm$ 0.07 \\
\hline
\end{tabular}
\caption{{Comparison of Ellipticity Estimates} for clusters in common with the
current
study, Kontizas et al. (1987), and Zepka~\& Dottori (1987). Frenk~\& Fall
(1982) and
Geisler~\& Hodge (1980) did not state what radii ellipticities were measured
at.
The column `Radius' lists the radii, in arcseconds, that
Kontizas et al. (1989) or Zepka~\& Dottori (1987)
estimated ellipticities over. `BDS' gives the ellipticities derived by the
current study for
the radii, while `K' and `ZD' list the ellipticity given by
Kontizas et al. (1989) or Zepka~\& Dottori (1987). The latter paper presented
two
different profiles for NGC 2214, which have been indicated as (a) and (b) in
`Range'.
\label{table:third_table} Kontizas et al. (1989) estimated an uncertainty of
0.03
for their ellipticity values.
}
\end{table}


\section{Simulations}

To test the reliability of the analysis
techniques, artificial elliptical star cluster images with known parameters
were generated.
This was to test if the input parameters could be successfully derived by the
two
methods.
The magnitude distribution was calculated using the IRAF {\it starlist}
function,
which allowed power, uniform, Salpeter model (a  best fit function to the data
of McCuskey (1966)), and Bahcall \& Soneira (1980) functions to be used.
It was decided to use an adjusted spherical
King (1962; 1966a,b) model for the placement of stars inside the cluster,
on the basis that spherical King models are
algebraically simple and have been widely applied to the LMC clusters
(e.g. Chun, 1978; Elson, 1991; Fischer { et al}, 1992, 1993). x and y
positions were generated using a pseudo-random number generator (which tests
revealed to have no bias). These values were then converted across to
polar coordinates centred on the (user set) artificial cluster centre.
The tidal, or limiting, radius of the cluster was assumed to vary elliptically.
Given the polar co-ordinates of the random position,
a tidal radius was generated for the point. If the point did
not fall within this limit, another random position was generated. Once a
point fell within its appropriate tidal limit, another random number between 0
and 1
was generated. This was compared with a radial distribution scaled
by the tidal radius at the polar angle $ \theta $, allowing ellipticity to be
included.
If the random number fell below the
distribution's probability at the point's radius, a star image was assumed to
exist. See Figure~\ref{figure:tres_bien} for a  King test distribution.
Finally, the elliptical distribution could be rotated
about its centre. This allowed two 5000 star image $ e $ = 0.5 King
distributions to be placed at right angles to each other.
Reduction of the
resulting image by the automatic ellipse fitting software was expected
to produce results of zero ellipticity.
A slightly greater ellipticity of 0.03 was achieved.
A higher ellipticity in the centre of this test image was due to the greater
effect of the positions of large
individual bright stars in the small fitting region, where they are also
more probable due to the increased stellar density towards the centre of a
cluster.
Above a radius of 3 full width half maxima reliable results were
being obtained.

\begin{figure}
\vspace{7.5cm}
\caption{{King Profile Distribution:} Star positions for a King distribution
0.5
ellipticity artificial cluster are shown.
\label{figure:tres_bien}
}
\end{figure}

As a further test, `flat' (uniform probability within the ellipse boundary)
distributions were sliced up in $ \theta $ (e.g. halves, eighths, etc.),
with each slice being cut into equal area segments, so as to check that the
radial
distribution was being scaled correctly with $ \theta $ (see Figure
\ref{figure:thingie}). The solid lines in this figure plot the stellar density
when
the cluster had been split into two halves containing 40 equal area segments
each.
The segments were considered to be sectors of a circle. Given that the
inner segment radius was 31.5 pixels, a
density of 124.4 stars was expected for
those segments inside a 100 pixel radius (the $ b $ axis). Examination
of the 18 segments within a 95 pixel radius gave a mean of 122.5 $ \pm $
10.1 stars per segment, and a median of 123.3. The mean density
over the 80 segments was 62.5 $ \pm $ 41.9, as expected. When the
distribution was sliced into sixths (the dotted lines), the density fell
by the expected third. Also note the dropoff of the two sectors centred
on the y axis is more rapid due to the smaller radial size of the sectors, and
that they have fallen off completely just above the $ b $ axis length
as could be expected for circular regions there.

Given the success of the flat distribution in this, and
that the King function used did approach the empirical inner and outer
radii functions of King (1966a), we are confident that the Monte-Carlo
placement of the stars was performed correctly.
Since the co-ordinates and magnitudes of the stars were
available, the IRAF {\it mkob} task was used to create artificial frames. A
Gaussian
profile was used for the Point Spread Function (PSF) of the 20,000 stars per
artificial frame,
together with a read noise value of 7.35 electrons and a
gain of 4.27 electrons per decimal unit (being the values for the TH7882
chip used to collect the real observations). 2.4'' seeing was assumed in the
tests, corresponding to the best seeing conditions experienced at MJUO by this
project. An aim
was to have a smooth distribution of faint stars, with a few bright ones
scattered
around in it.

\begin{figure}
\vspace{7.5cm}
\caption{Star counts in a sliced flat distribution:
5000 stars were randomly distributed in a flat elliptical
distribution with $ a $\label{figure:thingie}
= 200 and  $ e $ = 0.5.
}
\end{figure}

The first test image was of a 0.5 ellipticity cluster, with scaled King radii
based on
Chun's (1978) values for the old LMC cluster NGC 1835. These values were 500
and 15
pixels for the tidal and core radii respectively. These latter values were
chosen
for realism, and used in all the simulations presented below. Trials varying
the magnitude zero
point had no effect on the derived trends and results. The lowest threshold
value
used was barely above the background value, being at three standard deviations
of the noise
above the background. The derived position angle agreed well
with the input value, falling within 5 degrees of it. Ellipticity increased
linearly from an inner value of 0.3, reaching the input value at the outer
radii, and
then dramatically dropping at the last ``isophotes''. Such a low ellipticity
halo
containing an elliptical cluster was noted for NGC 2214 by Bhatia~\&
MacGillivray (1988),
although the cause in our simulation was simply that the cluster distribution
extended
off the right boundary of the image (due to slightly asymmetric centring of the
cluster in it). This
also resulted in the outer ellipses being skewed to the left (but not
vertically as symmetry
was maintained in that direction).

Concerned that the trends apparent in the image might be due to the placement
of bright
stars, whose brightness would skew ellipse fits, another image was created with
the
brightest three magnitudes excluded. 441 stars were `lost' by this process.
Fitting
showed that throughout the cluster the ellipticity of this smooth distribution
was
within 0.03 of the input ellipticity, until the boundary problem mentioned
above was
met. The position angle was stable, although systematically overestimated by
one degree,
while the x and y cluster centres were stable about the input values. Similar
trends
were found for simulations with all 20,000 stars set to the faint magnitude of
-1, although
they were slightly closer to the input values (as might be expected since the
combined
luminosity distribution gradients were more uniform across the image).  Such
results are in line with what could be expected for smooth distributions, such
as in
old LMC clusters where the bright stars have long since evolved.
Further trials using the original magnitude distribution (i.e. with the bright
stars),
but with different star placements
and ellipticities, showed the following:
\begin{enumerate}
\item The major axis angle could vary by $ \sim 80^{\rm o}$ with radius,
becoming more
variable as $ e $ approached zero. However, large variations of $\theta$ over
small radial
distances were found in a very ($ e \:
= \: 0.5 $) elliptical distribution, due to the random placement of bright
unresolved stars.
\item
No particular trend in ellipticity with radius appeared to be preferred.
Examples were
found where ellipticity increased, remained constant, or decreased with radius.
\item
Generally the input ellipticity was reached in the outer regions by the ellipse
fitting.
However, without
prior knowledge of this value, it would be difficult to determine that the
`real'
value of the cluster had been determined. Often, in the simulations it was only
met
for one or two points.
\end{enumerate}

However, such tests removing bright stars, are not very realistic as the bright
stars
are correctly `removed' from the cluster image. In practise, it is
common (see e.g. Fischer {\it et al}, 1992, 1993; Elson, 1991) to use a
PSF to subtract out the bright stars from an image. This is because it
has been widely recognized that such stars will bias ellipse (and profile
model) fits
by their placement and intensity. We therefore used the IRAF tasks {\it
daofind} and
{\it substar}, which are
based on DAOphot (Stetson, 1987). Outer radius bright stars in uncrowded
areas were used in an iterative process to construct the empirical corrections
to
a Gaussian PSF. This process concluded when tests showed that stars in
uncrowded regions
were being cleanly subtracted from the image. All the stars in the image that
could be
identified by DAOphot were,
and the bright ones (above a user set intensity limit) subtracted off using
the modeled PSF. The moment analysis software was then run on the image.
Ellipticity was
found to increase with radius, reaching the input value at $ \sim $ 4 times the
PSF's
FWHM, and then dropping away (before the outer isophotes were met). The range
of the
trend was 0.15 in $ e $, which is comparable with the trends found in our
observations
and Zepka \& Dottori (1987). The position angle was always within 3 degrees of
the input
value, excluding severe disturbance in the cluster centre. Similar results
can be seen in the results of Zepka \& Dottori (1987), contributing to their
gradients. These trends are
presumably due to the unresolved bright stars in the cluster centre. It is
interesting
to note that moment fitting of the original image showed the ellipticity was
recovered
more rapidly with radius than the star subtracted image, and that this
ellipticity was
more constant until the boundary problem was met. It is likely that poor
background
estimates in crowded regions leads to either positive (under-subtraction) or
negative
(over-subtraction) residuals,
resulting in spurious radial gradients.

{\it Ellipse} was also run on some of the test images, to see how well it
agreed with
the moments technique in light of our concerns mentioned above. Even in
clusters of 0.4
and 0.5 ellipticity trends were evident in $ e $, accompanied by $\theta$
ranging
up to 70 degrees. Low
ellipticities were found for cluster centres. The input ellipticity was not
recovered over substantial radial intervals
by the software, and was not obvious in plots of $e$ against radius (i.e
substantial plateaus corresponding to the input ellipticity's value were not
seen).
Problems were often encountered with {\it ellipse}
failing to iterate towards a solution. It does not appear that {\it ellipse} is
suitable
for young clusters, with their clumpy distributions, even when measures are
taken to
exclude the brightest pixels from given ellipse fits.


\section{Conclusions}

Radial variations in the ellipticity and position angle of LMC clusters have
previously
been taken to indicate that they are triaxial structures in equilibrium (Zepka
and
Dottori, 1987). We believe, in light of our simulations, that caution should be
exercised in the interpretation of ellipse fitting of young populous clusters.
Stochastic
effects in the placement of bright stars, which can not all be both detected
and
cleanly subtracted from an image using standard (and previously used)
techniques, will
result in spurious trends. Such problems will not be evident in large bodies
observed
at low resolution (such as M81) or evolved old clusters (such as NGC 1835),
both with smooth radial distributions. The fact that trends were observed in
NGC 1835
suggests that its radial variations may be real.


\section{Acknowledgements}

The authors are grateful for generous time allocations at the
Mount John University Observatory,
and to the New Zealand Foundation for Research,
Science and Technology for partial funding of this project, in
conjunction with the VUW Internal Research Grant Committee.
TB acknowledges partial support during this study by the
inaugural R.H.T Bates Postgraduate Scholarship of the Royal Society
of New Zealand. We would like to thank Acorn New Zealand for the
loan of an R260 machine, on which some of this paper's work was
performed, and also acknowledge the help of the Referee, Dr. R.S. Stobie.


\bsp
\label{lastpage}
\end{document}